\documentclass[12pt]{article}
\usepackage{amssymb,latexsym}
\usepackage[mathscr]{eucal}

\newcommand{\be}{\begin{equation}}
\newcommand{\ee}{\end{equation}}
\newcommand{\bea}{\begin{eqnarray}}
\newcommand{\eea}{\end{eqnarray}}

\begin{document}
%%%%%%%%%%%%%%%%%%%%%%%  FRONTESPIZIO  %%%%%%%%%%%%%%%%%%%%%%%%%%%%%%%%%%%
\begin{titlepage}
\topmargin 1.5cm

%%%%%%%%%%%%%%%%%%%%%%%  TITOLO  %%%%%%%%%%%%%%%%%%%%%%%%%%%%%%%%%%%%%%%%%
\begin{center} {\large \sc Instantons and the 5D
$U(1)$ gauge theory
\\with extra
adjoint} \\

\vspace{0.5cm}
%%%%%%%%%%%%%%%%%%%%%%%%  AUTORI  %%%%%%%%%%%%%%%%%%%%%%%%%%%%%%%%%%%%%%%%%
{\sc Rubik Poghossian }\\
{\sl Yerevan Physics Institute,\\
Alikhanian Br. st. 2, 0036 Yerevan, Armenia}\\
e-mail: poghos@yerphi.am \\
\vskip 0.2cm
{\sc Marine Samsonyan}\\
{\sl Dipartimento di Fisica, Universit\'a di Roma ``Tor Vergata''\\
I.N.F.N. Sezione di Roma II,\\
Via della Ricerca Scientifica, 00133 Roma, Italy}\\
e-mail: marine.samsonian@roma2.infn.it \\
\vskip 0.2cm
\end{center}
\vskip 0.5cm
%%%%%%%%%%%%%%%%%%%%%%%%  ABSTRACT  %%%%%%%%%%%%%%%%%%%%%%%%%%%%%%%%%%%%%%
\begin{center}
{\large \bf Abstract}
\end{center}
{\hspace{0.5 cm} In this paper we compute the partition function
of 5D supersymmetric $U(1)$ gauge theory with extra adjoint matter
in general $\Omega$-background. It is well known that such
partition functions encode very rich topological information. We
show in particular that unlike the case with no extra matter, the
partition function with extra adjoint at some special values of
the parameters directly reproduces the generating function for the
Poincare polynomial of the moduli space of instantons. Comparing
our results with those recently obtained by Iqbal et. al., who
used the refined topological vertex method, we present our
comments on apparent discrepancies.}

\par    \vfill
\end{titlepage}
%%%%%%%%%%%%%%%%%%%%%%%%%%%%%%%%%%%%%%%%%%%%%%%%%%%%%%%%%%%%%%%%%%%%%%%%%%
%%%%%%%%%%%%%%%%%%%%%%  INIZIO TESTO  %%%%%%%%%%%%%%%%%%%%%%%%%%%%%%%%%%%%
%%%%%%%%%%%%%%%%%%%%%%%%%%%%%%%%%%%%%%%%%%%%%%%%%%%%%%%%%%%%%%%%%%%%%%%%%%
\addtolength{\baselineskip}{0.3\baselineskip}
%%%%%%%%%%%%%%%%%%%%%%%%%%%%%%%%%%%%%%%%%%%%%%%%%%%%%%%%%%%%%%%%%%%%%%%%%%
%%%%%%%%%%%%%%%%%%%%%%%%%%%%%%%%%%%%%%%%%%%%%%%%%%%%%%%%%%%%%%%%%%%%%%%%%%
%%%%%%%%%%%%%%%%%%%%%%%%%%%%%%%%%%%%%%%%%%%%%%%%%%%%%%%%%%%%%%%%%%%%%%%%%%
%%%%%%%%%%%%%%%%%%%%%%%%%%%%%%%%%%%%%%%%%%%%%%%%%%%%%%%%%%%%%%%%%%%%%%%%%%
%%%%%%%%%%%%%%%%%%%%%%  SECTION 1     %%%%%%%%%%%%%%%%%%%%%%%%%%%%%%%%%%%%
%%%%%%%%%%%%%%%%%%%%%%%%%%%%%%%%%%%%%%%%%%%%%%%%%%%%%%%%%%%%%%%%%%%%%%%%%%

\setcounter{section}{0}
\section{Introduction}
\setcounter{equation}{0} \hspace{0.5 cm}Recent progress in
understanding non-perturbative phenomena in supersymmetric
Yang-Mills theories due to direct multi-instanton calculations is
quite impressive. Two main ideas played essential role in all this
developments. First was the realization that the Supersymmetric
Yang-Mills action induced to the moduli space of instantons can be
represented in terms of closed, equivariant with respect to the
diagonal part of the gauge group, forms \cite{FPS}. This
observation leads to a crucial simplifications reducing SYM path
integral to an integral over the stable with respect to the action
of the diagonal part of the gauge group subset of the moduli space
of instantons. The next brilliant idea, which is the corner stone
for all further developments was suggested by Nekrasov in
\cite{Nek}. The idea is to generalize the theory involving into
the game in equal setting besides the already mentioned global
diagonal gauge transformations also the diagonal part of the
(Euclidean) space-time rotations. Why this is so crucial because
the subset of the instanton moduli space invariant under this
combined group action appears to consist only of finite number of
points.

In the case of the gauge group $U(N)$ this fixed points set is in
one to one correspondence with the set of array of Young diagrams
$\vec{Y}=(Y_1,...,Y_N)$ with total number of boxes $|\vec{Y}|$
being equal to the instanton charge $k$. Thus, to calculate path
integral for the various "protected" by super-symmetry physical
quantities one needs to know only the pattern how the combined
group acts in the neighborhoods of the fixed points. All this
information can be encoded in the character of the group action in
the tangent space at given fixed points. An elegant formula for
this character which played a significant role in both physical
and mathematical applications was proposed in \cite{FP} (see eq.
(\ref{char})). Let us note at once that combining space time
rotations with gauge transformations besides giving huge
computational advantage due to finiteness of the fixed point set,
has also a major physical significance generalizing the theory to
the case with certain nontrivial graviphoton backgrounds
\cite{Nek}. In order to recover the standard flat space quantities
(say the Seiberg-Witten prepotential of ${\cal N}=2$
super-Yang-Mills theory) one should take the limit when the space
time rotation angles vanish. It is shown by Nekrasov and Okounkov
\cite{Nek_Okoun} that in this limit the sum over the arrays of
Young diagrams is dominated by a single array with specific
"limiting shape". This enables one to handle in this limit the
entire instanton sum exressing all relevant quantities in terms of
emerging Seiberg-Witten curve. This is essential since only the
entire sum and not its truncated part exhibits remarkable modular
properties which allows one to investigate rich phase structure of
SYM theories. This is why all the attempts to investigate the
instanton sums also in general case seems quite natural.
Unfortunately there was little progress till now in this direction
besides the simplest case of gage group $U(1)$. Though the $U(1)$
4D theory in flat background is trivial, the general 5D $U(1)$
theory compactified on a circle\footnote{Roughly speaking the main
technical difference between 4D and 5D cases is that in the former
case the above mentioned combined group enters into the game in
the infinitesimal level while in the latter case the main role is
played by finite group elements.} being rather nontrivial
nevertheless in many cases admits full solution. In what follows
we investigate the partition function of 5D gauge theory with an
extra adjoint hypermultiplet. It is not surprising that such
partition functions encode very rich topological information. As a
manifestation we argue that unlike the case with no extra matter,
at some special values of the parameters this partition function
directly reproduces the generating function of the Poincare
polynomial for the moduli space of instantons. We check this
conclusion explicitly computing the partition function in the case
of gauge group $U(1)$. We compare our result with that of recently
obtained by Iqbal et. al. \cite{Iqbal} who used the refined
topological vertex method to find the same partition function and
present our comments on discrepancies we found.

\section{The $U(1)$ theory with adjoint matter}
\setcounter{equation}{0} The weight decomposition of the torus
action on the tangent space at the fixed point
$\vec{Y}=(Y_1,\dots,Y_N)$ is given by \cite{FP} \bea \chi=
\sum_{\alpha ,\beta =1}^N e_\beta e_\alpha^{-1} \left\{\sum_{s\in
Y_\alpha}\left(T_1^{-l_\beta(s)}T_2^{a_\alpha(s)+1}\right)+\sum_{s\in
Y_\beta}\left(T_1^{l_\alpha(s)+1}T_2^{-a_\beta(s)}\right)\right\},\label{char}
\eea where $e_1,...,e_N$ are elements of (complexified) maximal
torus of the gauge group $U(N)$ and $T_1, T_2$ belong to the
maximal torus of the (Euclidean) space-time rotations, $a_\alpha
(s)$ ($l_\alpha (s)$) measures the distance from the location of
the box $s$ to the edge of the young diagram $Y_\alpha$ in
vertical (horizontal) direction.

The 5D partition function in the pure ${\cal N}=2$ theory could be
read off from above character \bea Z=\sum_{\vec{Y}}\frac{
\mathbf{q}^{|\vec{Y}|}}{\prod_{\alpha ,\beta =1}^N \prod_{s\in
Y_\alpha}\left(1-e_\beta e_\alpha^{-1}
T_1^{-l_\beta(s)}T_2^{a_\alpha(s)+1}\right)\prod_{s\in
Y_\beta}\left(1-e_\beta
e_\alpha^{-1}T_1^{l_\alpha(s)+1}T_2^{-a_\beta(s)}\right)}\,\,\,
\eea From the mathematical point of view this quantity could be
regarded as the character of torus action on the space of
holomorphic functions of the moduli space of instantons. The
Nekrasov's partition function for 4D theory could be obtained
tuning the parameters $\mathbf{q}\rightarrow \beta^{2N}
\mathbf{q}$, $T_1\rightarrow \exp-\beta \epsilon_1 $,
$T_2\rightarrow \exp-\beta \epsilon_2 $, $e_\alpha \rightarrow
-\beta v_\alpha$ and tending $\beta \rightarrow 0$, where
$v_1,...,v_N$ are the expectation values of the chiral superfield
and $\epsilon_1$, $\epsilon_2$ characterize the strength of the
graviphoton background (sometimes called $\Omega$-background).

Fortunately instanton counting is powerful enough to handle also
the cases when an extra hypermultipet in adjoint or several
fundamental hypermultiplets are present. In the case with adjoint
hypermultiplet instead of (\ref{char}) one starts with the (super)
character \cite{fucito} \bea \chi=(1-T_m) \sum_{\alpha ,\beta
=1}^N e_\beta e_\alpha^{-1} \left\{\sum_{s\in
Y_\alpha}\left(T_1^{-l_\beta(s)}T_2^{a_\alpha(s)+1}\right)+\sum_{s\in
Y_\beta}\left(T_1^{l_\alpha(s)+1}T_2^{-a_\beta(s)}\right)\right\}.\label{char_adj}
\eea One way to interpret this character is to imagine that each
(complex) 1d eigenspace of the torus action is complemented by a
grassmanian eigenspace with exactly the same eigenvalues of the
torus action. In addition an extra $U(1)$ action is introduced so
that $T_m\in U(1)$ acts trivially on bosonic directions while
acting on each grassmanian coordinate in its fundamental
representation. Then (\ref{char_adj}) is the super-trace of the
extended torus action on the super-tangent space at given fixed
point. The corresponding 5D partition function now reads: \bea
Z=\sum_{\vec{Y}}\mathbf{q}^{|Y|}\prod_{\alpha ,\beta =1}^N
\prod_{s\in Y_\alpha}\frac{\left(1-T_m e_\beta e_\alpha^{-1}
T_1^{-l_\beta(s)}T_2^{a_\alpha(s)+1}\right)}{\left(1-e_\beta
e_\alpha^{-1}
T_1^{-l_\beta(s)}T_2^{a_\alpha(s)+1}\right)}\prod_{s\in
Y_\beta}\frac{\left(1-T_m e_\beta
e_\alpha^{-1}T_1^{l_\alpha(s)+1}T_2^{-a_\beta(s)}\right)}{\left(1-e_\beta
e_\alpha^{-1}T_1^{l_\alpha(s)+1}T_2^{-a_\beta(s)}\right)}\nonumber
\\ \label{Zadj}\eea
Each term here could be thought as trace over the space of local
holomorphic forms, with parameter $T_m$ counting the degrees of
forms. Hence the sum over the fixed points is expected to give the
super-trace over the globally defined holomorphic forms. We see
that $Z_{adj}$ is an extremely rich quantity from both physical
and mathematical point of view. It is interesting to note that at
special values of the parameters $Z_{adj}$ directly reproduces the
generating function for the Poincare polynomial of the moduli
space of $U(N)$ instantons. Indeed, following \cite{Nak_lect} part
(3.3) let us assume that $T_2\gg T_{a_1}>\cdots >T_{a_N}\gg
T_1>0$. It is easy to see that in the limit when all these
parameters go to zero each fraction under the products in
(\ref{char_adj}) tends to $T_m$ or $1$ depending whether we have a
negative weight direction or not (see the classification of
negative directions in \cite{Nak_lect}, proof of corollary 3.10).
We will see this explicitly in the simplest case $N=1$ when the
moduli space of instantons coincides with the Hilbert scheme of
points on $\mathbb{C}^2$.

From now on we will restrict ourselves to the simplest case of
$U(1)$ gauge group, when the partition function could be computed
in a closed way. The partition function of the pure ${\cal N}=2$
$U(1)$ theory has the form \cite{Nak_blowup} \bea
Z=\sum_{Y}\frac{\mathbf{q}^{|Y|}}{\prod_{s\in Y}\left(1-
T_1^{-l(s)}T_2^{a(s)+1}\right)\left(1-T_1^{l(s)+1}T_2^{-a(s)}\right)}
\nonumber \\=\exp \left(\sum_{n=1}^\infty \frac{\mathbf{q}^n}{n
(1-T_1^n)(1-T_2^n)}\right). \eea This remarkable combinatorial
identity in the 4D limit and in "self dual" case
$\epsilon_1=-\epsilon_2$ boils down to the Burnside's theorem \bea
\sum_{|\lambda|=n}(dim \,\mathbf{R}_\lambda )^2=n!, \eea where
$\mathbf{R}_\lambda$ is the irreducible representation of the
symmetric group given by the Young diagram $\lambda$.

Now let us turn to the $U(1)$ theory with adjoint matter. Doing
low instanton calculations using (\ref{Zadj}) is straightforward
and gives \bea \log Z_{adj}=\frac{\mathbf{q}(1+T_m
\mathbf{q}+T_m^2 \mathbf{q}^2+T_m^3 \mathbf{q}^3)
(1-T_m T_1)(1-T_m T_2)}{(1- T_1)(1-T_2)}+\nonumber \\
\frac{\mathbf{q}^2(1+T_m^2\mathbf{q}^2) (1-T_m^2 T_1^2)(1-T_m^2
T_2^2)}{2(1- T_1^2)(1-T_2^2)}+\frac{\mathbf{q}^3(1-T_m^3
T_1^3)(1-T_m^3 T_2^3)}{3(1- T_1^3)(1-T_2^3)}\nonumber
\\+\frac{\mathbf{q}^4 (1-T_m^4 T_1^4)(1-T_m^4 T_2^4)}{4(1-
T_1^4)(1-T_2^4)}+\textit{O}(\mathbf{q}^4). \eea These drove us to
the conjecture that the exact formula is \bea \log
Z_{adj}=\sum_{n=1}^\infty\frac{\mathbf{q}^n (1-T_m^n
T_1^n)(1-T_m^n T_2^n)}{n(1-
T_1^n)(1-T_2^n)(1-T_m^n\mathbf{q}^n)},\label{log_Z_adj} \eea which
is equivalent to the highly nontrivial combinatorial identity \bea
Z_{adj}=\sum_{Y}\mathbf{q}^{|Y|}\prod_{s\in Y}\frac{\left(1-
T_mT_1^{-l(s)}T_2^{a(s)+1}\right)\left(1-T_mT_1^{l(s)+1}T_2^{-a(s)}\right)}{
\left(1-T_1^{-l(s)}T_2^{a(s)+1}\right)\left(1-T_1^{l(s)+1}T_2^{-a(s)}\right)}
\\
=\exp \left(\sum_{n=1}^\infty
\frac{\mathbf{q}^n(1-(T_mT_1)^n)(1-(T_mT_2)^n)}{n
(1-T_1^n)(1-T_2^n)(1-(T_m\mathbf{q})^n)}\right). \eea

 Indeed calculations with Mathematica code up to 10
instantons further convinced us that this formula is indeed
correct. As a further check let us go to the limit when
$T_1\rightarrow 0$, $T_2\rightarrow 0$. As we have explained above
one expects to find the generating function of Poincare polynomial
for Hilbert scheme of points on $\mathbb{C}^2$. An easy
calculation yields: \bea Z_{adj}|_{T_1,T_2=0}=\exp
\sum_{n=1}^\infty \frac{\mathbf{q}^n}{n(1-T^n_m
\mathbf{q}^n)}=\nonumber \\
\exp\sum_{n=1}^\infty \sum_{k=0}^\infty
\frac{(\mathbf{q}^{1+k}T_m^k)^n}{n}=\prod_{k=0}^\infty\frac{1}{1-T_m^k\mathbf{q}^{k+1}},\eea
which indeed after identifying $T_m$ with Poincare parameter $t^2$
reproduces the well known result (see e.g. \cite{Nak_book}). Now
let us go back to the general case. In various domains of the
variables $T_1$, $T_2$ in similar manner we can represent
(\ref{log_Z_adj}) as infinite product. Let us consider separately
the cases: \\
(a) $|T_1|<1$, $|T_2|<1$, $|T_m \mathbf{q}|<1$\\
In this region (\ref{log_Z_adj}) could be rewritten as \bea
Z_{adj}=\exp\left\{\sum_{n=1}^\infty \sum_{k,i,j=0}^\infty
\frac{\mathbf{q}^n}{n}T_1^{ni}T_2^{nj}(T_m
\mathbf{q})^{nk}(1-T_m^n T_1^n)(1-T_m^n T_2^n)\right\}. \eea
Performing summation over $n$ first we get \bea
Z_{adj}=\prod_{i,j,k=0}^\infty
\frac{(1-\mathbf{q}^{k+1}T_m^{k+1}T_1^{i+1}T_2^{j})
(1-\mathbf{q}^{k+1}T_m^{k+1}T_1^{i}T_2^{j+1})}{(1-\mathbf{q}^{k+1}T_m^{k}T_1^{i}T_2^{j})
(1-\mathbf{q}^{k+1}T_m^{k+2}T_1^{i+1}T_2^{j+1})}.
\label{Z_prod_a}\eea
(b) $|T_1|>1$, $|T_2|<1$, $|T_m \mathbf{q}|<1$\\
In this region we expand (\ref{log_Z_adj}) over $1/T_1$: \bea
Z_{adj}=\exp\left\{\sum_{n=1}^\infty \sum_{k,i,j=0}^\infty
\frac{-\mathbf{q}^n}{n}T_1^{-ni}T_2^{nj}(T_m
\mathbf{q})^{nk}(1-T_m^n T_1^n)(1-T_m^n T_2^n)T_1^{-n}\right\},
\eea which leads to \bea Z_{adj}=\prod_{i,j,k=0}^\infty
\frac{(1-\mathbf{q}^{k+1}T_m^{k}T_1^{-i-1}T_2^{j})
(1-\mathbf{q}^{k+1}T_m^{k+2}T_1^{-i}T_2^{j+1})}{(1-\mathbf{q}^{k+1}T_m^{k+1}T_1^{-i}T_2^{j})
(1-\mathbf{q}^{k+1}T_m^{k+1}T_1^{-i-1}T_2^{j+1})}.\label{Z_prod_b}\eea

Recently Iqbal, Koz\c{c}az and Shabir \cite{Iqbal} have computed
the partition function of these $U(1)$ adjoint theory using the
refined topological vertex formalizm \cite{Iqbal_Vafa}. And, since
the formula (\ref{log_Z_adj}) was known to the present authors for
quite a while, we performed a detailed comparison of these
results. To make contact with the formulae of Iqbal et. al. we
need the following dictionary: $T_m=Q_m(t/q)^{1/2}$, $T_1=1/t$,
$T_2=q$, $\mathbf{q}=Q(q/t)^{1/2}$. In terms of these variables
the
equations (\ref{Z_prod_a}) and (\ref{Z_prod_b}) take the form: \\
(a) $|t|>1$, $|q|<1$, $|QQ_m|<1$\\
\bea Z_{adj}=\prod_{i,j,k=1}^\infty
\frac{(1-Q^{k}Q_m^{k}q^{i-1}t^{-j})
(1-Q^{k}Q_m^{k}q^{i}t^{1-j})}{(1-Q^{k}Q_m^{k+1}
q^{i-\frac{1}{2}}t^{-j+\frac{1}{2}})(1-Q^{k}Q_m^{k-1}q^{i-\frac{1}{2}}t^{-j+\frac{1}{2}})},
\eea and \\
(b) $|t|<1$, $|q|<1$, $|QQ_m|<1$\\
\bea Z_{adj}=\prod_{i,j,k=1}^\infty
\frac{(1-Q^{k}Q_m^{k+1}q^{i-\frac{1}{2}}t^{j-\frac{1}{2}})
(1-Q^{k}Q_m^{k-1}q^{i-\frac{1}{2}}t^{j-\frac{1}{2}})}{(1-Q^{k}Q_m^{k}q^{i-1}t^{j-1})
(1-Q^{k}Q_m^{k}q^{i}t^{j})}.\eea These equations come rather
close, but certainly do not coincide with those given in
\cite{Iqbal} at the end of the part 3.2. The reason for this
discrepancy seems to us as follows. According to \cite{Iqbal} the
refined topological vertex method for the 5D $U(1)$ theory with
adjoint matter leads to (see eq. (4.6) of \cite{Iqbal}; below we
omit the "perturbative part" \\$\prod_{i',j'=1}^\infty
(1-Q_mq^{-\rho_{i'}}t^{-\rho_{j'}})$ ) \bea Z=\prod_{k=1}^\infty
(1-Q^kQ_m^k)^{-1} \prod_{i,j=1}^\infty
(1-Q^kQ_m^{k-1}q^{-\rho_i}t^{-\rho_j})\\
(1-Q^kQ_m^{k}q^{\rho_i-1/2}t^{-\rho_j+1/2})
(1-Q^kQ_m^{k}q^{-\rho_i+1/2}t^{\rho_j-1/2})
(1-Q^kQ_m^{k+1}q^{\rho_i}t^{\rho_j}), \nonumber \label{Z_RTV}\eea
where $\rho_i=-i+1/2$. But four factors under the product over
$i,j$ have different, excluding each other regions of convergence.
Thus this infinite product should be treated very carefully.
Unfortunately the authors of \cite{Iqbal} do not tell what
analytic continuation procedure they have adopted to pass from
their eq. (4.6) to those presented at the end of the part 3.2, but
we will demonstrate now that one, perhaps the simplest approach
directly leads to our conjectural formula (\ref{log_Z_adj}). We
simply examine the product over each factor separately within its
region of convergence and only after that continue analytically to
a common region of the parameters. Thus for the first factor in
(\ref{Z_RTV}) we have \bea \prod_{k=1}^\infty
(1-Q^kQ_m^k)^{-1}=\exp \sum_{n,k=1}^\infty
\frac{(QQ_m)^{nk}}{n}=\exp \sum_{n=1}^\infty
\frac{(QQ_m)^{n}}{n(1-(QQ_m)^n)}. \eea For the next factor
(assuming $q<1$, $t<1$) \bea \prod_{k,i,j=1}^\infty
(1-Q^kQ_m^{k-1}q^{i-\frac{1}{2}}t^{j-\frac{1}{2}})=\exp
\sum_{n,k,i,j=1}^\infty
\frac{-Q^{kn}Q_m^{(k-1)n}q^{(i-\frac{1}{2})n}t^{(j-\frac{1}{2})n}}{n}
\nonumber \\
=\exp \sum_{n=1}^\infty
\frac{-Q^nq^{\frac{n}{2}}t^{\frac{n}{2}}}{(1-(QQ_m)^n)
(1-q^n)(1-t^n)}.\eea Similarly for $q>1$, $t<1$ \bea
\prod_{k,i,j=1}^\infty (1-Q^kQ_m^{k}q^{-i}t^{j})=\exp
\sum_{n=1}^\infty \frac{-Q^nQ_m^{n}q^{-n}t^{n}}{n(1-(QQ_m)^n)
(1-q^{-n})(1-t^n)}, \eea for $q<1$, $t>1$ \bea
\prod_{k,i,j=1}^\infty (1-Q^kQ_m^{k}q^{i}t^{-j})=\exp
\sum_{n=1}^\infty \frac{-Q^nQ_m^{n}q^{n}t^{-n}}{n(1-(QQ_m)^n)
(1-q^{n})(1-t^{-n})}, \eea and, finally for $q>1$, $t>1$ \bea
\prod_{k,i,j=1}^\infty
(1-Q^kQ_m^{k+1}q^{-i+\frac{1}{2}}t^{-j+\frac{1}{2}})=\exp
\sum_{n=1}^\infty
\frac{-Q^nQ_m^{2n}q^{-\frac{n}{2}}t^{-\frac{n}{2}}}{n(1-(QQ_m)^n)
(1-q^{-n})(1-t^{-n})}. \eea Note that the r.h.s.'s of above
expressions are defined also outside of their initial convergence
region. Combining all these together we get \bea  Z=\exp
\sum_{n=1}^\infty
\frac{(QQ_m)^n(q^{\frac{n}{2}}t^{\frac{n}{2}}-Q_m^n)
(q^{\frac{n}{2}}t^{\frac{n}{2}}-Q_m^{-n})}{n
(1-(QQ_m)^n)(1-q^n)(1-t^n)}, \eea which in terms of the parameters
$\mathbf{q}$, $T_1$, $T_2$ exactly coincides with our conjectural
result (\ref{log_Z_adj}).

\section*{Acknowledgements}
It is a pleasure to thank R.Flume, F.Fucito and J.-F.Morales for
discussions. R.P. would like to thank I.N.F.N. for supporting a
visit to the University of Rome II, "Tor Vergata". This work was
supported by the Institutional Partnership grant of Humboldt
foundation of Germany. M.S. acknowledges support of RTN grant
MRTN-CT-2004-503369.

\end{document}